# Harnessing a Vibroacoustic Mode for Enabling Smart Functions on Surface Acoustic Wave Devices - Application to Icing Monitoring and Deicing

Atefeh Karimzadeh[1]*, Uhland Weissker[1], Jaime del Moral[2], Andreas Winkler[1], Ana Borrás[2], Agustin R. González-Elipe[2], Stefan Jacob[3]*

[1] Leibniz IFW-Dresden, Helmholtzstr. 20, 01069 Dresden, Germany.

[2] Nanotechnology on Surfaces and Plasma Lab, Materials Science Institute of Seville (CSIC-Univ. Sevilla), Americo Vespucio 49, Sevilla, 41092 Spain.

[3] Physikalisch-Technische Bundesanstalt (PTB), Bundesallee 100, 38116 Braunschweig, Germany.

E-mail: a.karimzadeh.66@gmail.com, a.karimzadeharani@ifw-dresden.de; stefan.jacob@ptb.de.

**Abstract**

Microacoustic wave devices are essential components in the RF electronics and MEMS industry with increasing impact in various sensing and actuation applications. Reliable and smart operation of acoustic wave devices at low costs would cause a crucial advancement. Herein, we present the enablement of temperature and mechanical sensing capabilities in a Rayleigh-mode standing surface acoustic wave (sSAW) chip device by harnessing an acoustic shear plate wave mode using the same set of electrodes. Most importantly, this mode is excited by switching the polarity of the sSAW transducer electrodes by simple electronics, allowing for direct and inexpensive compatibility with an existing setup. We validated the method in the emergent topic of surface de-icing by continuously monitoring temperature and water liquid-solid phase changes using the plate wave mode, and on-demand Rayleigh-wave deicing with a negligible energy cost. The flexibility for adapting the system to different scenarios, loads and scalability opens the path to impact in lab-on-a-chip, IoT technology, and sectors requiring autonomous acoustic wave actuators.

**Keywords: Automated microacoustic device; shear plate wave sensor; surface acoustic wave actuator; sustainable design; smart de-icing; icing monitoring.**

**Introduction**

During recent years, Surface Acoustic Waves (SAW) devices have been exploited by research and industry in numerous nano- and macro-scale applications, e.g. in biology, robotics, electronics, printing, and surface cleaning [1-6]. However, complying with the industry 4.0 paradigm as well as with recent demands of lab-on-a-chip and IoT technologies requires the implementation of highly sensitive, reliable, flexible, and low-cost solutions enabling both sensing and actuation within a single device. Currently, those functions are integrated by altering the device or by adding new components, which often increases the system's cost, size, and complexity [6-9].

A conventional SAW device, as presented in Fig. 1a (left), consists of a piezoelectric substrate with comb-shaped interdigital transducer electrodes (IDTs). For standing SAW excitation as shown here, the transducers are pairwise excited by a radio frequency signal in the range of a few tenths of MHz to several GHz depending on the specific application and the IDTs design. When activated at the IDTs operation frequency, elastic Rayleigh-mode acoustic waves with nano-scale amplitude propagate along the surface of the substrate and resonate between the IDT pairs, carrying energy that can be used for actuation, e.g. for liquid nebulization or, as it has been recently demonstrate, for surface deicing [10, 11]. In actuation applications, resonance peaks can be several hundred kHz to a few MHz in width, providing reproducible and stable activation conditions. Those setups can also be used in sensing applications, for example in delay line configurations [12]. In some cases, e.g. temperature measurements, microacoustic sensing often requires a narrow-band resonance peak to realize accurate monitoring. For this case, acoustic plate waves (PW) of Lamb and shear types are preferential, as they show narrow-band resonances with a bandwidth of a few kHz or less that provide excellent sensing capabilities [13-16]. Usually, PW activation is induced with pairs of extended electrodes, which are different in size and shape from the IDTs used in SAW devices, making the cointegration of standing surface acoustic waves (sSAW) actuation and PW sensing in a single device difficult.

Addressing this problem, here we demonstrate that switching the polarity of the IDT electrodes of conventional sSAW devices, as shown in Fig. 1a (right), could excite a strong PW thickness shear mode in the low MHz frequency range that can be used for surface sensing. Switching between both acoustic modes is realized by a simple relay circuit inserted between the RF signal generator and the device, i.e., without introducing any conceptual changes on existing setup designs, and by using only the existing IDT electrodes. From now on, we will refer to PW operation (sensing) and sSAW operation (actuation), depending on the acoustic wave mode that is excited in the device.

To prove the concept in a realistic scenario, we present the development and operation of the first smart acoustic de-icing system integrating ultrafast ice detection by PW and reliable ice melting by Rayleigh sSAW. Ice accretion on surfaces is a critical issue for daily life as well as for industrial and large-scale applications [17-19]. During the last few years, important advances have been made in both active (e.g. removal of accreted ice by heating or using anti-icing fluids) and passive (surface engineering treatments to reduce ice accretion and adhesion) approaches [16, 19-22]. However, enhancing energy efficiency, reducing environmental impact, and increasing compatibility with industrial-relevant systems are challenges that still need smart and reliable solutions. In this context, de-icing by acoustic wave may become an enabling technology if these early results demonstrating low-power consumption, ice detection capability, and synergy with anti-icing coatings are exploitable in a smart and autonomously operated mode, i.e. by monitoring the formation and volume of accreted ice and automatically activating the deicing using the least required power, as shown in this work.

**Results and discussion:**

The sSAW chip employed in this study has been recently used for successful deicing and anti-icing applications (see Jacob, et al 2023 [20] for extended details). Fig. 1 illustrates how the PW mode enabled the smart operation of the acoustic deicing setup. A thickness-shear PW mode can be activated by properly poling the IDTs as shown in Fig. 1a (right), i.e. in an attempt to mimic the nature of extended electrodes. This acoustic mode depends on the plate (chip device) thickness and was found to be located at a frequency between 3.5 and 3.6 MHz in most of the tested cases. We used the magnitude of the electric reflection coefficient ($|S_{11}|$) in PW operation as a suitable sensor signal characterized by a sharp drop at the shear-thickness PW resonance frequency with a width of less than 1 kHz (Fig. 1 b). The $S_{11}$ of the PW mode was tracked using a network analyzer. $S_{11}$ measurements can be run in a low-power state with electrical power of less than one mW. The PW resonance was sensitive to any parameter that affected the substrate's stiffness, such as variations in temperature or pressure on the substrate surface.

For testing the PW/sSAW hybrid mode of operation, we first dispensed a water droplet with a diameter of 15 mm on the surface, covering about 15% of the open surface of the chip and mimicking a rain droplet on a sensor surface. Subsequently, we cooled the device from 300 K to 263 K in the controlled environment of a climate chamber. Fig. 1b demonstrates the ability of a low-power excited PW mode to log the substrate's temperature through the resonance frequency. The resonance frequency of the substrate in PW operation, which was obtained at the minimum of the measured $S_{11}$, increased significantly as the environmental temperature decreased, while the shape of the curve only slightly changed, provided that liquid water remained in a supercooled state. We were also able to detect the water droplet freezing on the chip surface which was connected to a sudden change in the resonance's amplitude. As shown in Fig. 1b, changing the water droplet phase to ice caused a remarkable drop in the intensity of the $S_{11}$ curve and a shift in resonance frequency, in accordance with earlier findings in del Moral et al. 2023 [16] using extended electrodes for the excitation of PWs. These results underline the great potential of the herein disclosed PW mode for sensing and recording a variety of processes in sSAW devices initially designed for sSAW activation.

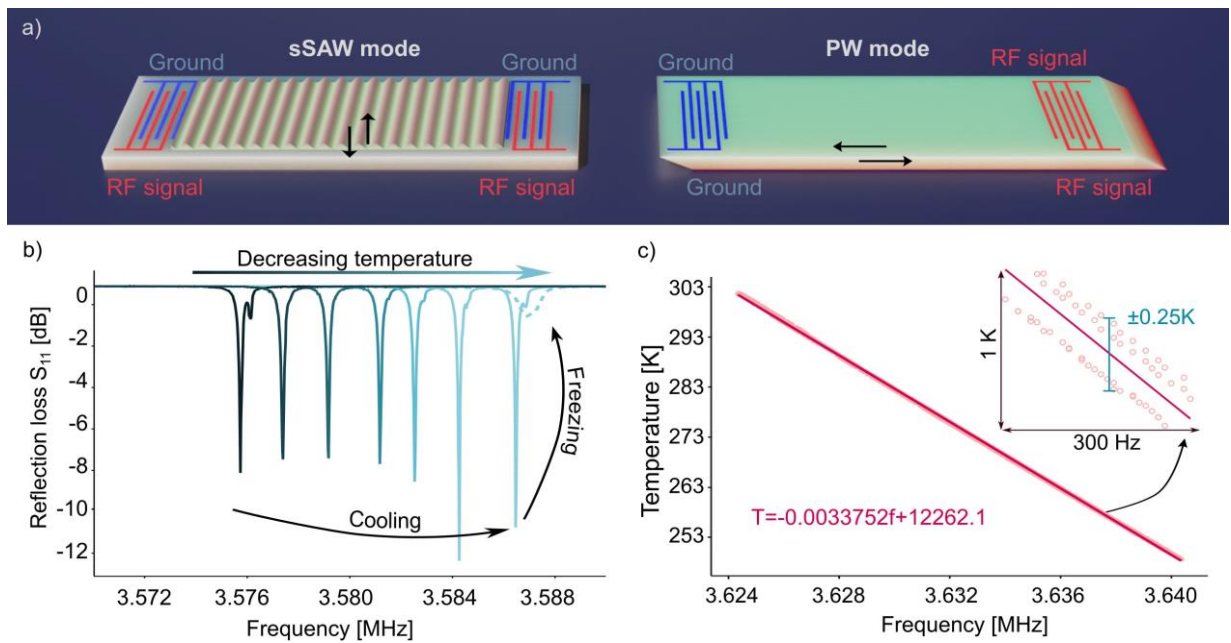

Fig. 1. a) Schematic illustration of the device operated in sSAW mode (left) and PW mode (right). The colors of the electrodes indicate their potential (red for RF signal and blue for ground); b) Evolution of $S_{11}$ during operation of the PW sensor for sensing temperature and water-ice phase change. The temperature is indicated by the color of the spectra; c) Relationship between the temperature of the substrate and the frequency of the shear-thickness PW's resonance, suggesting a linear correlation with a low derivation.

Second, we determined the rate of change of the PW mode's resonance frequency over the temperature. A K-type thermocouple was mounted to the backside of the substrate to measure its temperature. The thermocouple was isolated from the surrounding air using a 2 mm thick isolation tape. The substrate's temperature measured by the thermocouple was then correlated to the resonance frequency of the PW mode. The temperature and the PW mode's resonance frequency were tracked automatically every 20 s for a time of 19 h, cycling the temperature variation between 250 K and 300 K for three times. The relation between temperature and the PW mode's resonance frequency was close to linear within this temperature range. The results suggest that the PW mode in sSAW devices could be calibrated during fabrication for temperature measurements, using a simple linear regression and a few calibrated temperature points. In our case, the maximum deviation between temperatures measured at the same resonance frequency was about 0.5 K, resulting in a maximum error between the linear regression and the predicted temperature of ± 0.25 K (Fig. 1c). A temperature-sensing accuracy of less than 0.25 K is sufficient for most sSAW applications in printing, surface cleaning and atomizing [20, 23, 24], making additional devices for temperature tracking, e.g. thermocouples, redundant.

Third, we showed that the PW sensing can be used to create automated actuation, e.g., for smart de-icing (see Fig. 2, Fig. 3, and Video S1). A water droplet was left on the chip surface at subzero temperature inside the climate chamber with a surrounding temperature of -15 deg, and an unsupervised computer algorithm constantly switched between PW operation and sSAW operation modes to melt the ice and keep the droplet in liquid form. The simple algorithm (see Fig. 3b) acquired the PW resonance frequency every 2 s, extracting the temperature and the state of the water through the recording and analysis of the $S_{11}$ spectra (Fig. 2a), while a video of the water/ice transformation was simultaneously recorded (steps i) to iv) in Fig. 2c and Video S1). Once freezing occurred (Fig. 2c ii)), the significant and immediate change of the resonance frequency's amplitude (Fig. 2a ii) triggered the program to switch to sSAW operation. The corresponding $S_{11}$ curve recorded with the network analyzer at 1 mW is presented in Fig. 2b. Then, a power smaller than 7 W was applied at 33.2 MHz to efficiently melt the frozen droplet (Figs. 2c iii)). By intermittent switching to PW operation during deicing, the algorithm detected complete melting (Fig. 2c iv)) and stopped the excitation, guaranteeing an energy-efficient on-demand activation protocol.

A perfect correlation was apparent between the state of water/ice as directly deduced from the video recording (snapshots i-iv) in Fig. 2c) and the $S_{11}$ curve profiles in Fig. 2a. A video of the automated routine is available in the S1. The time that was required for monitoring the state of the water/ice droplet intermittently during deicing depended on the logic circuit used and the setup of the $S_{11}$ tracking device. In the presented experiment, the use of relay switches in combination with a programmable microcontroller unit (MCU) and a commercial general-purpose network analyzer resulted in monitoring times shorter than a second. Although sufficient for deicing, this time could be drastically reduced if faster logics (e.g. MEMS circuits) and more advanced tracking algorithms were used for time-critical applications. Furthermore, since only the magnitude of $S_{11}$ (and no phase information) was used by the described program, future simplifications can be envisaged for the sensing electronics.

Adding sensing capabilities through PW modes can enable sSAW chips to function for smart applications at low costs and with little or no redesign on the existing layouts. In the reported experiment, an unsupervised deicing on a large transparent surface was demonstrated. However, the suggested method is not restricted to that application. A shear-thickness PW mode primarily depends on the thickness of the substrate and not on other system parameters. Therefore, it would be reproduced on a variety of sSAW chips, which are manufactured on substrates with the same thickness, and might be used in other fields of application.

We hypothesize that this shear-thickness PW mode has been so far overlooked and therefore not exploited for sensing along with sSAW actuators because of the narrow band-width of the PW resonance and its appearance outside the typical sSAW frequency range. We hope that this brief report contributes to that other researches may identify and harness the PW mode to extend the capabilities of technologies based on sSAW actuators, especially in sensitive and bio-related applications. The suggested method for sensing is sensitive to any physical changes that cause variation of the substrate stiffness, e.g. temperature, pressure, or change of surface material due to a chemical reaction, and could be applied to any sSAW devices with a range of resonance frequencies higher than the resonance frequency of the PW mode. The field is ample and susceptible for extensive research to examine and improve the PW mode sensing performance in combination with alternative sSAW actuator designs (e.g., including structured and coated substrates) and/or to improve the electronics needed for switching and sensing using simplified circuits.

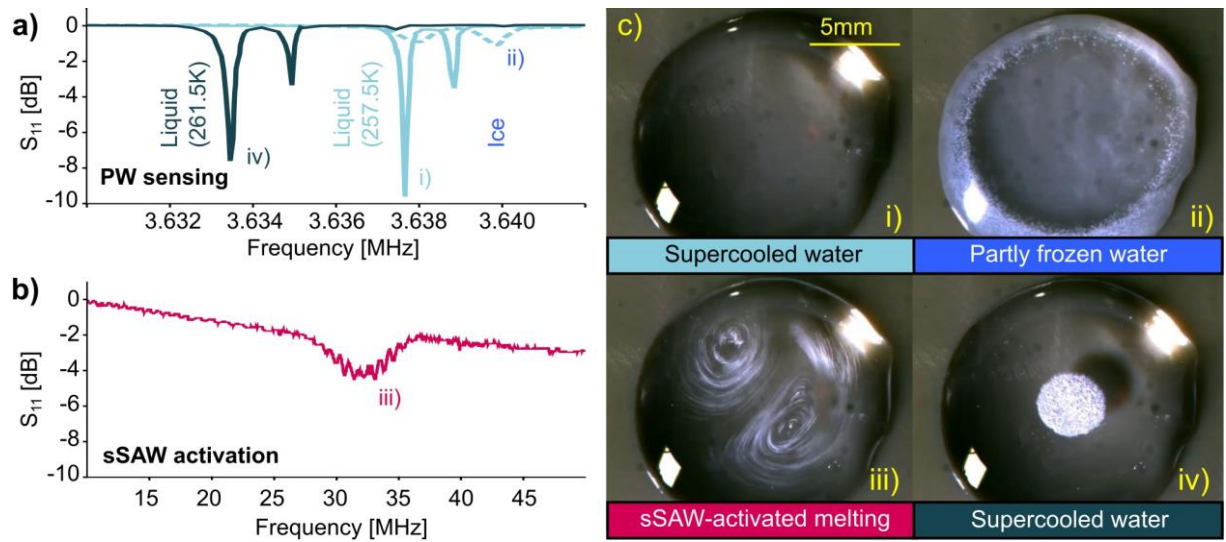

Fig. 2. Smart operation of a SAW device for deicing applications: a) $S_{11}$ spectra recorded during PW sensing the temperature and state of water/ice droplet at the different stages of the freezing/melting process, b) $S_{11}$ spectrum corresponding to the sSAW mode use to melt the ice (see also Video S1), c) Snapshots of the water/ice droplet: i) supercooled droplet of water; ii) Mixture water/ice during melting; iii) partially melted ice during sSAW excitation; iv) Water droplet after melting through the application of the sSAW. The inclusion in the center corresponds to small ice island floating on the water.

**Methods:**

SAW-based deicing actuator chips with dimensions of 70 mm x 35 mm were prepared from a 4" black double-side polished $LiNbO_3$ 128°YX wafer (Hangzhou Freqcontrol Electronic Technology Ltd., China) with an average thickness of 525 μm as the substrate. Comb-shape IDTs formed by a thin film consisting of 5 nm thick Ti and 295 nm Al were prepared using magnetron sputtering and lift-off photolithography. The whole actuator surface (except the IDTs pads) was covered by a 100 nm thick $SiO_2$ protection layer to prevent the electro-corrosion of Al IDTs. The IDTs were designed for a Rayleigh-type sSAW activation at a frequency of 32.5 MHz, covering only a small area at the sides of the chip (less than 1.5% of the chip area). Details about the IDT structure design and the production process of the device have been described in our previous publication [20]. The same electrodes were used to excite the substrate with a shear PW mode at a resonance frequency of 3.576 MHz at room temperature by changing the IDTs polarity as shown in Fig. 1a (right).

RF frequency was applied to the IDTs using a network analyzer or a specialized SAW signal source (BSG F20, BelektroniG GmbH, Germany) through a printed circuit board (PCB) with spring pins contacted with the IDTs pads. The same device was used for receiving and analyzing the $S_{11}$ curve. A combination of relay (HF-Relay Omron G6K-2F-RF 5V) and microcontroller (Pocket Science Lab) as shown in Fig. 3a, was used to switch between sSAW actuation and PW sensing operation. The switching scheme was dictated by a simple program that followed the steps in the pseudocode in Fig. 3b. The switching (line 4 and line 7 in Fig. 3b) was realized by turning the 5V input signal to the relays on and off. The ice detection algorithm (line 6 in Fig. 3b) compared the amplitude from the current $S_{11}$ measurement to the previous measurement, exploiting the distinct and instantaneous drop of the $S_{11}$ curve after freezing (Fig. 2a i) to ii)). The $S_{11}$ measurement and the excitation algorithms (line 5 and line 9 in Fig. 3b) interfaced pre-defined functions in the network analyzer. Excitation and sensing times (line 13 and line 10 in Fig. 3b) and powers (line 3 and line 8 in Fig. 3b) can be adjusted based on the desired application.

The performance of the deicing actuator device was tested in a climate chamber (Binder MKF-56, BINDER GmbH, Germany). A droplet of distilled water was placed at the center of the device using a syringe. Then, it was cooled down by reducing the temperature of the climate chamber until the droplet phase changed to ice. A USB camera was placed inside the environmental chamber and normal to the surface of the device to optically record the droplet phase change during the experiment.

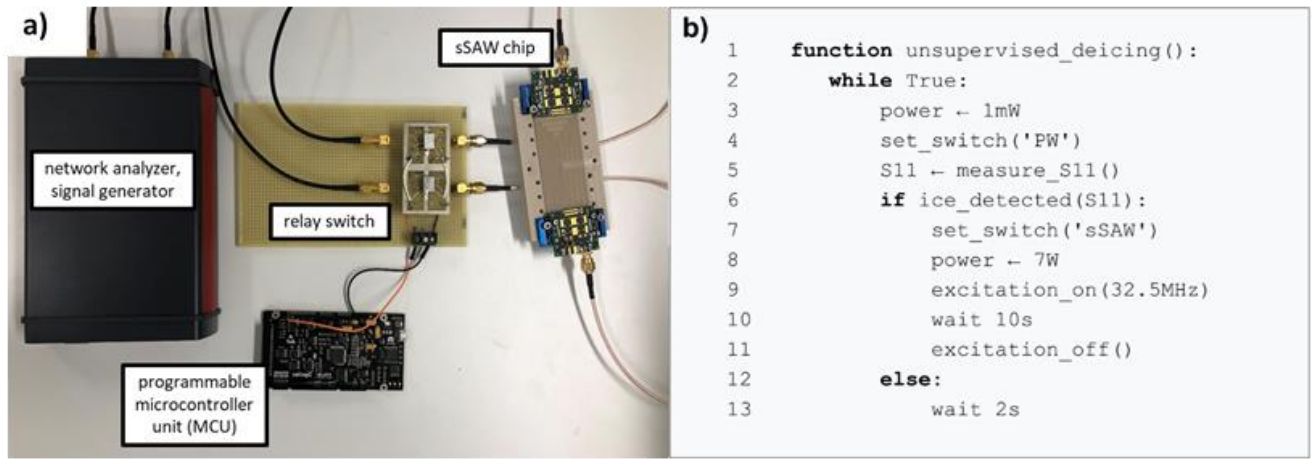

**Fig. 3.** a) Smart deicing setup including the SAW device and the relay circuit, b) sensing and deicing algorithm as pseudocode.

**Acknowledgements**

This article is part of a project funded by the EU H2020 program under grant agreement 899352 (FETOPEN-01-2018-2019-2020 – SOUNDofICE). The authors gratefully acknowledge Steve Wohlrab from Leibniz-Institut für Festkörper- und Werkstoffforschung (IFW) Dresden for their support during sample preparation.